\documentclass[twocolumn, prb, showpacs, superscriptaddress]{revtex4-1}

\usepackage{graphicx}
\usepackage{amsmath}
\usepackage{hyperref}

\newcommand{\metal}{\ensuremath{\mathcal{M}}}
\newcommand{\bond}[1]{{\sffamily\bfseries #1}}
\newcommand{\mof}{MOF-74}
\newcommand{\reaction}{H$_2$O~$\rightarrow$~OH+H}

\graphicspath{{./figures/}}
\newcommand{\UTD}{Department of Materials Science and Engineering,
University of Texas at Dallas, Richardson, Texas 75080, USA}
\newcommand{\WFU}{Department of Physics, Wake Forest University,
Winston-Salem, North Carolina 27109, USA} \newcommand{\RU}{Department of
Chemistry and Chemical Biology, Rutgers University, Piscataway, New
Jersey 08854, USA}

\begin{document}

\title{Understanding
and Controlling Water Stability of MOF-74}

\author{S. Zuluaga}                  \affiliation{\WFU}
\author{E. M. A. Fuentes-Fernandez}  \affiliation{\UTD}
\author{K. Tan}                      \affiliation{\UTD}
\author{F. Xu}                       \affiliation{\RU}
\author{J. Li}                       \affiliation{\RU}
\author{Y. J. Chabal}                \affiliation{\UTD}
\author{T. Thonhauser}\email{thonhauser@wfu.edu}\affiliation{\WFU}

\date{\today}

\begin{abstract}
Metal organic
framework (MOF) materials in general, and MOF-74 in particular, have
promising properties for many technologically important processes.
However, their instability under humid conditions severely restricts
practical use. We show that this instability and the accompanying
reduction of the CO$_2$ uptake capacity of MOF-74 under humid conditions
originate in the water dissociation reaction H$_2$O~$\rightarrow$~OH+H
at the metal centers. After this dissociation, the OH groups coordinate
to the metal centers, explaining the reduction in the MOF's CO$_2$
uptake capacity. This reduction thus strongly depends on the catalytic
activity of MOF-74 towards the water dissociation reaction. We further
show that---while the water molecules themselves only have a negligible
effect on the crystal structure of MOF-74---the OH and H products of the
dissociation reaction significantly weaken the MOF framework and lead to
the observed crystal structure breakdown. With this knowledge, we
propose a way to suppress this particular reaction by modifying the
{\mof} structure to increase the water dissociation energy barrier and
thus control the stability of the system under humid conditions.
\end{abstract}
\maketitle

\section{Introduction}
\label{intro}

Metal organic framework (MOF) materials have promising properties
towards technologically important applications such as gas storage and
sequestration,\cite{Liu_2012:progress_adsorption-based,
Murray_2009:hydrogen_storage, Li_2011:carbon_dioxide,
Qiu_2009:molecular_engineering, Nijem_2012:tuning_gate,
Lee_2015:small-molecule_adsorption, Zhao_2008:current_status,
Rosi_2003:hydrogen_storage, Wu_2012:commensurate_adsorption,
He_2014:methane_storage}
sensing,\cite{Kreno_2012:metal-organic_framework,
Serre_2007:role_solvent-host, Allendorf_2008:stress-induced_chemical,
Tan_2011:mechanical_properties, Hu_2014:luminescent_metal-organic}
polymerization, \cite{Uemura_2009:polymerization_reactions,
Vitorino_2009:lanthanide_metal} luminescence,
\cite{Allendorf_2009:luminescent_metal,
White_2009:near-infrared_luminescent} non-linear optics,
\cite{Bordiga_2004:electronic_vibrational} magnetic networks,
\cite{Kurmoo_2009:magnetic_metal-organic} targeted drug delivery,
\cite{Horcajada_2010:porous_metal-organic-framework} multiferroics,
\cite{Stroppa_2011:electric_control, Stroppa_2013:hybrid_improper,
Di-Sante_2013:tuning_ferroelectric} and
catalysis.\cite{Wu_2007:heterogeneous_asymmetric,
Lee_2009:metal-organic_framework, Zou_2006:preparation_adsorption,
Luz_2010:bridging_homogeneous} In particular, MOF-74 [\metal$_2$(dobdc),
\metal\ = Mg$^{2+}$, Zn$^{2+}$, Ni$^{2+}$, Co$^{2+}$, and
dobdc=2,5-dihydroxybenzenedicarboxylic acid] exhibits unsaturated metal
centers ideal for the adsorption of small molecules such as
H$_2$,\cite{Liu_2008:increasing_density, Zhou_2008:enhanced_h2}
CO$_2$,\cite{Wu_2010:adsorption_sites,
Dietzel_2008:adsorption_properties, Caskey_2008:dramatic_tuning}
N$_2$,\cite{Valenzano_2010:computational_experimental} and
CH$_4$,\cite{Wu_2009:high-capacity_methane} among others. Unfortunately,
the presence of water vapor can reduce the gas uptake significantly and
can even destroy the crystal structure of several members of the MOF-74
family,\cite{Kizzie_2011:effect_humidity, Remy_2013:selective_dynamic,
Liu_2011:stability_effects, Schoenecker_2012:effect_water,
DeCoste_2013:effect_water} restricting its practical use. The exact
mechanisms responsible for this structural instability and the reduction
in the CO$_2$ gas uptake are poorly understood and their relation is
unclear. For example: After Mg-MOF-74 is exposed to humid conditions,
its CO$_2$ gas uptake capacity is severely reduced, while its crystal
structure remains intact.\cite{Liu_2011:stability_effects} On the other
hand, Ni-MOF-74 suffers a smaller reduction in CO$_2$ uptake capacity,
but its crystal structure is considerable damaged by water
vapor.\cite{Liu_2011:stability_effects}

A great effort has been made to enhance the stability of MOFs under
humid conditions.\cite{Burtch_2014:water_stability,
Li_2013:water_thermally, Tan_2012:stability_hydrolyzation,
Han_2013:mof_stability} For example, Gao and co-workers obtained two new
rht-MOFs (rht-MOF-triazolate and rht-MOF-pyrazolate) that are
isostructural with rht-MOF-1, but exhibit an enhancement in their
stability under humid conditions.\cite{Gao_2015:remote_stabilization}
Decoste and co-workers have been able to enhance the stability of Cu-BTC
(1,3,5 benzenetricarboxylic acid, BTC) MOF by treating it with
plasma-enhanced chemical vapor deposition of perfluorohexane, which
results in a hydrophobic form of Cu-BTC with a remarkable stability
under humid conditions.\cite{Decoste_2012:enhanced_stability} Other
groups have shown that it is possible to enhance the stability of three
micro porous Zn-MOFs by introducing water repellent groups such as
methyl, shielding the metal centers from water molecules and thus
enhancing the stability of the MOF under humid
conditions.\cite{Ma_2011:tuning_moisture} Concerning MOF-74, Jiao and
co-workers have synthesized and characterized a series of MM-MOF-74
structures, replacing some of the Mg metal centers of Mg-{\mof} with either Ni or Co
atoms.\cite{Jiao_2015:tuning_kinetic} Their results show that the
addition of a small amount of Ni (16~mol\%) remarkably enhances the
water stability of those systems.

As mentioned before, it is not fully understood how the water molecules
affect the crystal structure and the adsorption properties of the {\mof}
system.  However, some preliminary work has been done on this topic,
which inspired our own research.  Decoste and
co-workers\cite{DeCoste_2013:effect_water} hypothesize that the
adsorption of water on the metal centers of Mg-MOF-74 breaks the bonds
that bind the one-dimensional channels together and thus hinders the
diffusion of small molecules to the interior of the MOF. As a results,
the gas uptake capacity is reduced, but leaves the structure of the
channels intact, explaining why x-ray patterns do not show significant
changes in the crystal structure of this MOF. Han {\it et
al.}\cite{Han_2010:molecular_dynamics} performed molecular dynamics
simulations on the adsorption of water on Zn-MOF-74, finding that before
it looses its crystal structure, the water molecules dissociate into OH
and H at the metal centers. However, further experimental and
theoretical work is needed to truly understand this interesting effect.

Only recently, our group has found experimental proof of the
dissociation of water at the metal centers of MOF-74 at temperatures
above 150~$^{\circ}$C,\cite{Tan_2014:water_reaction,
Tan_2015:water_interactions} where we found first hints concerning the
H$_2$O~$\rightarrow$~OH+H reaction and its relation to the CO$_2$ uptake
reduction. In the present work, combining experimental and {\it ab
initio} results, we uncover the exact mechanism that is responsible for
the degradation of MOF-74 under humid condition and elucidate the
connection between its structural instability and CO$_2$ uptake
decrease. We prove four points concerning the interaction of water with
the {\mof} structure: (i) Once the {\reaction} reaction takes place, the
OH poisons the metal centers of {\mof} and reduces the CO$_2$ gas uptake
capacity of the system; (ii) H$_2$O itself has only a negligible effect
on the linker--metal bonds---{\it{i.e}}.\ oxygen--metal
(O--$\mathcal{M}$), where we consider $\mathcal{M}=$ Zn, Mg, Ni, and
Co---and the crystal structure of {\mof}. The reaction products OH and H
are the ones responsible for the elongation of the O--$\mathcal{M}$
bonds and the breakdown of the crystal structure; (iii) different metal
centers have different O--$\mathcal{M}$ elongations in the presence of
OH and H groups. O--Mg and O--Co bonds are the least affected, while
O--Zn and O--Ni are the most affected; and finally, (iv) the {\reaction}
reaction can be suppressed by an appropriate linker modification.

\section{Results and Discussion}

\subsection{Fingerprint of the \reaction\ Reaction}

\begin{figure}[t]
\includegraphics[width=\columnwidth]{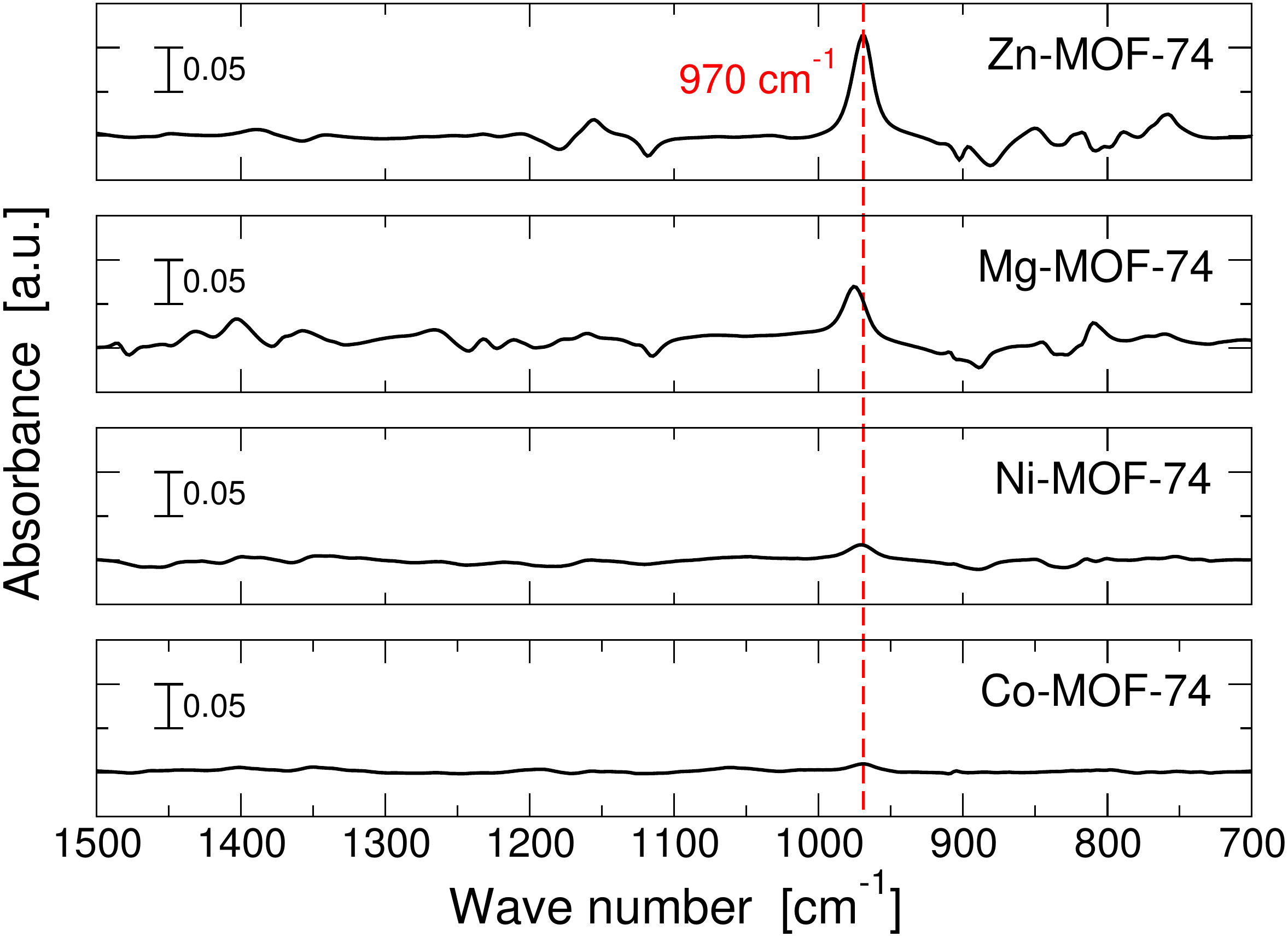}
\caption{\label{diff-IR_fig} IR absorption spectra of Zn-, Mg-, Ni-, and
Co-MOF-74 hydrated by introducing 8~Torr of D$_2$O vapor for 20 minutes
at 200~$^{\circ}$C, referenced to the activated MOF in vacuum. See the
supplementary information of the work of Tan and
co-workers\cite{Tan_2014:water_reaction} for more information.}
\end{figure}

Until recently, experiments have not been able to detect the
H$_2$O~$\rightarrow$~OH+H reaction in
MOF-74.\cite{Schoenecker_2012:effect_water, DeCoste_2013:effect_water,
Kizzie_2011:effect_humidity, Remy_2013:selective_dynamic,
Liu_2011:stability_effects} Our group was the first to detect its
fingerprint in form of a sharp peak in the IR spectrum of Zn-\mof\ at
970~cm$^{-1}$, which can only be observed when the MOF is exposed to
D$_2$O above 150~${^\circ}$C; for H$_2$O, this mode is at
1316~cm$^{-1}$, where it strongly couples to and gets drowned by modes
of the linker, making it impossible to
detect.\cite{Tan_2014:water_reaction, Tan_2015:water_interactions} For
this reason, we conducted our experiments in the present study with
D$_2$O instead of H$_2$O.  Although, for simplicity, throughout the text
we may still use the word ``water'' or ``H$_2$O.'' Furthermore, for Mg-,
Ni-, and Co-{\mof}, the peak is observed at 977~cm$^{-1}$,
971~cm$^{-1}$, and 969~cm$^{-1}$ (see Fig.~\ref{diff-IR_fig}), but we
will refer to all these peaks as the ``970~cm$^{-1}$ peak.''

\begin{figure}[t]
\centering
\includegraphics[width=0.8\columnwidth]{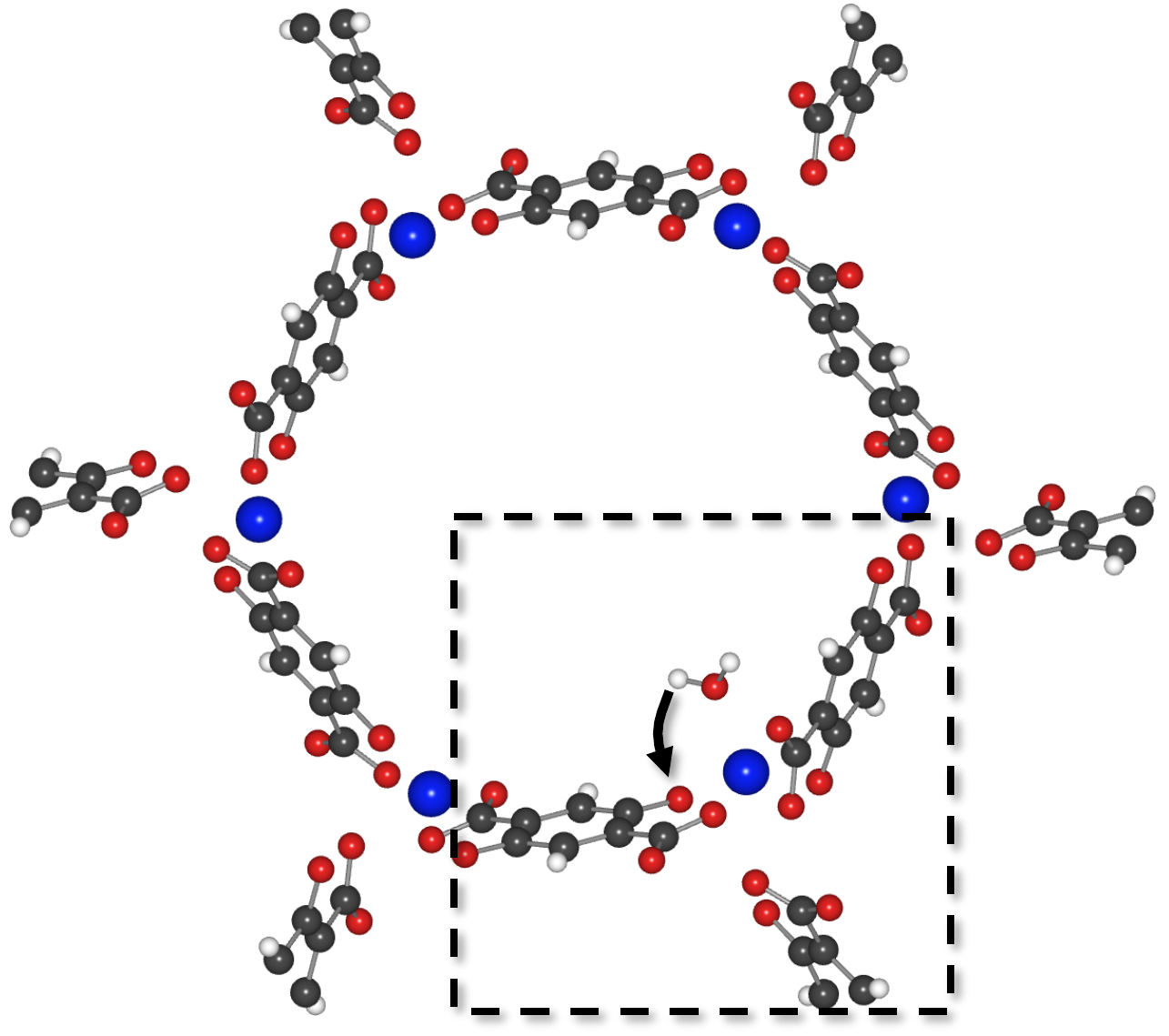}
\caption{\label{1H2O_fig} Graphical representation of MOF-74 and the
H$_2$O~$\rightarrow$~OH+H reaction. The hexagonal channel structure of
MOF-74 with its six equivalent open metal sites per unit cell is clearly
visible. The arrow indicates how the H atom of the water molecule is
transferred to the O atom of the linker. Black, red, white, and blue
spheres represent C, O, H, and metal atoms. The box indicates the
portion of MOF-74 visible in Fig.~\ref{linker-names_fig}, albeit from a
slightly different angle.}
\end{figure}

Using \emph{ab initio} simulations, we found the
H$_2$O~$\rightarrow$~OH+H reaction to proceed along the following
pathway: Once the water molecule is adsorbed at the metal center, one of
its H atoms is donated to the nearest O of the linker, as depicted in
Fig.~\ref{1H2O_fig}; the activation barrier for this process is
1.02~eV,\cite{Tan_2014:water_reaction} as discussed in detail below.

\subsection{Relation Between Amount of Water Dissociated and CO$_2$
Uptake Reduction}

\begin{figure}[t]
\includegraphics[width=1.0\columnwidth]{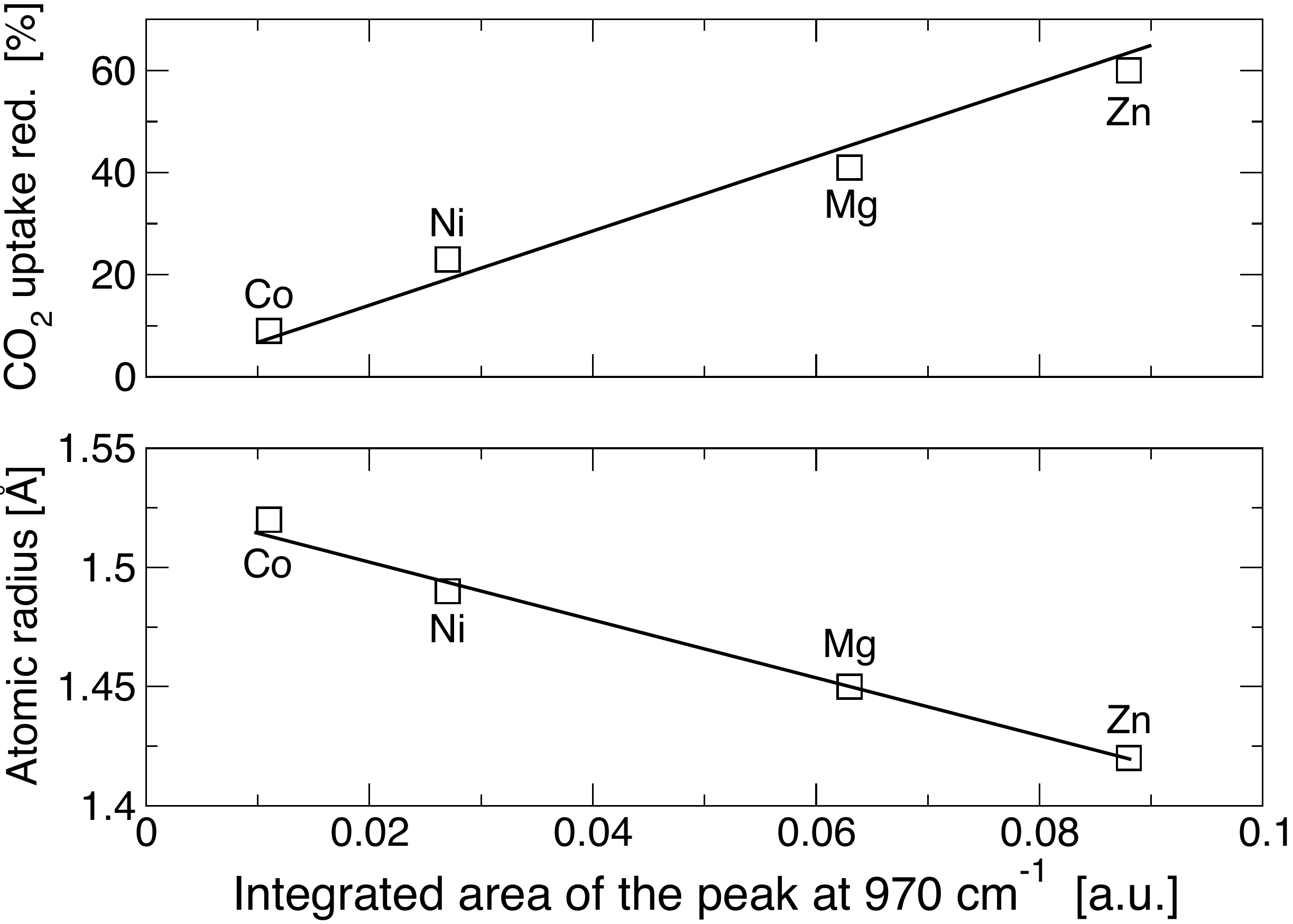}
\caption{\label{linear_fig}(top) Reduction in CO$_2$ uptake as a
function of the integrated area $\mathcal{A}$ of the peak in the IR
spectrum at 970~cm$^{-1}$ in Fig.~\ref{diff-IR_fig}, which is a measure
of the amount of water that has dissociated. The reduction is measured
after $\mathcal{M}$-MOF-74 ($\mathcal{M}=$ Zn, Mg, Ni, and Co) has been
exposed to 8~Torr of D$_2$O vapor at 200~$^{\circ}$C for 20 minutes.
Data taken from Table~\ref{stabl-descrpt_tab}. (bottom) The radius of
the metal centers (Zn, Mg, Ni and
Co)\cite{Clementi_1963:atomic_screening} vs.\ catalytic activity towards
the {\reaction} reaction (integrated area $\mathcal{A}$ under the peak
at 970~cm$^{-1}$). The straight lines are linear fits to the data.}
\end{figure}

We now analyze the relationship between the integrated area
$\mathcal{A}$ of the peak at 970~cm$^{-1}$, which is a direct measure of
the amount of water dissociated and the decrease in CO$_2$ uptake. In
Table~\ref{stabl-descrpt_tab} we show the integrated area of the peak at
970~cm$^{-1}$ and the CO$_2$ uptake reduction after the
$\mathcal{M}$-MOF-74 system ($\mathcal{M}=$ Zn, Mg, Ni, and Co) has been
exposed to 8~Torr of D$_2$O vapor at 100~$^{\circ}$C and
200~$^{\circ}$C. The dissociation reaction starts at
150~$^\circ$C,\cite{Tan_2014:water_reaction} so that the reported values
for 100~$^{\circ}$C and 200~$^{\circ}$C are clearly below and above the
threshold temperature. The table confirms that a significant reduction
in the CO$_2$ uptake only appears at elevated temperatures, {\it{i.e}}.\
once the water dissociation reaction has started. The table also reveals
the close relation between the \emph{amount} of water dissociated at the
metal centers ($\mathcal{A}$) and the reduction in CO$_2$ uptake. In
fact, the relationship is almost perfectly linear, as can bee seen in
the top panel of Fig.~\ref{linear_fig}. It follows that the dissociation
reaction is the key factor controlling the reduction in gas uptake when
MOF-74 is exposed to humid conditions: The metal centers of the MOF-74
are being poisoned by the OH groups after the dissociation reaction
takes place and are not available for binding CO$_2$ anymore---or any
other small molecules, for that matter.

\begin{table}[t]
\caption{\label{stabl-descrpt_tab} Integrated area $\mathcal{A}$ of the
peak at 970~cm$^{-1}$ in the IR spectrum and CO$_2$ uptake reduction
[\%] after $\mathcal{M}$-MOF-74 ($\mathcal{M}=$ Zn, Mg, Ni, and Co) has
been exposed to 8~Torr of D$_2$O vapor at 100~$^{\circ}$C and
200~$^{\circ}$C for 20 minutes. This data is also plotted in the top
panel of Fig.~\ref{linear_fig}, where its linear relationship becomes
obvious. At 100~$^\circ$C $\mathcal{A}$ falls below our level of
resolution and we report it as zero.}

\begin{tabular*}{\columnwidth}{@{\extracolsep{\fill}}lcccr@{}}\hline\hline
{$\mathcal{M}$}   &  \multicolumn{2}{c}{100~$^{\circ}$C}   &
\multicolumn{2}{c}{200~$^{\circ}$C} \\\cline{2-3}\cline{4-5}
& CO$_2$ red.  & $\mathcal{A}$        & CO$_2$ red.  & $\mathcal{A}$    \\\hline
Zn  &8\%  & 0.00  &60\%  &0.09\\
Mg  &7\%  & 0.00  &41\%  &0.06\\
Ni  &2\%  & 0.00  &23\%  &0.03\\
Co  &2\%  & 0.00  &9\%   &0.01\\\hline\hline
\end{tabular*}
\end{table}

As evident from Table~\ref{stabl-descrpt_tab} and Fig.~\ref{linear_fig},
MOFs with different metals show a varying degree of catalytic activity
towards the {\reaction} reaction. But, what causes and/or determines the
catalytic activity towards the water dissociation reaction for a
particular metal? Clearly, the catalytic activity is not related to how
strong the water binds to the metal
centers.\cite{Lee_2015:small-molecule_adsorption,
Canepa_2013:high-throughput_screening} However, if we plot the catalytic
activity (area $\mathcal{A}$ under the peak at 970~cm$^{-1}$) against
the calculated radius of the atoms,\cite{Clementi_1963:atomic_screening}
we obtain a perfectly linear fit, as can be seen in the lower panel of
Fig~\ref{linear_fig}. The reason for this is the following: in the
{\reaction} reaction, the metal centers interact with the O atom of the
water, while one of the H is transferred to the linker, see
Fig~\ref{1H2O_fig}. In other words, the water molecule interacts with
the metal center and the linker at the same time. As the size of the
metal centers decreases, the H atoms (of the water molecule) get closer
to the O atoms of the linker, increasing the interaction between them
and facilitating the transfer of the H to the linker. We thus can use
the size of the metal atoms as a predictor of the catalytic activity of
the {\mof} system towards the {\reaction} reaction.

\subsection{Relation to Structural Stability}

\begin{figure}[t]
\includegraphics[width=\columnwidth]{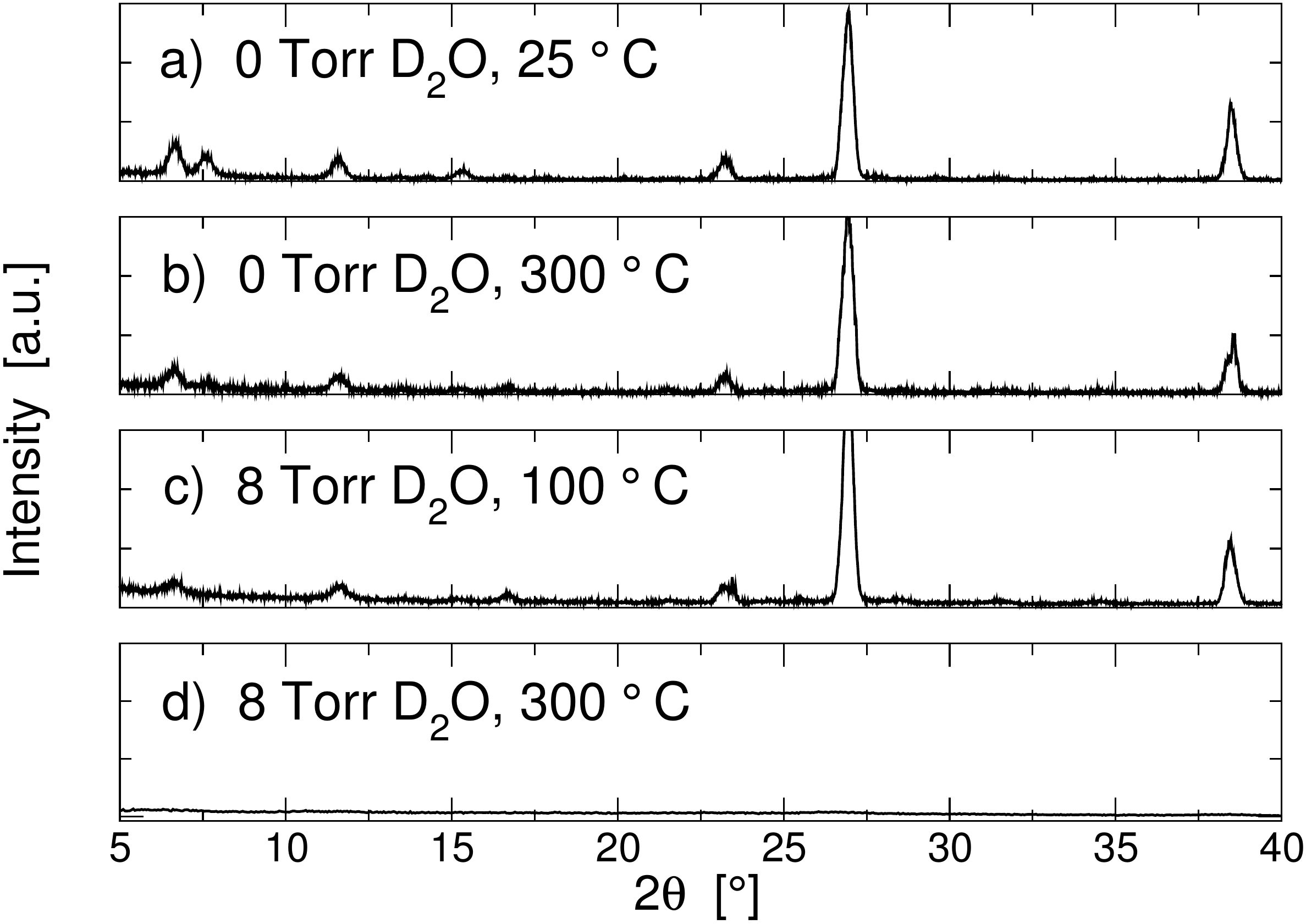}
\caption{\label{Zn_XRD_fig}(a) XRD patterns of the pristine Zn-{\mof}
before exposure to water at 25~$^{\circ}$C. (b) XRD patterns after
exposure to 300~$^{\circ}$C for 48 hours under vacuum (less than
20~mTorr). (c) XRD patterns after exposure to 100~$^{\circ}$C and 8~Torr
of D$_2$O for 48 hours. (d) XRD patterns after exposure to
300~$^{\circ}$C and 8~Torr of D$_2$O for 48 hours.}
\end{figure}

According to Table~\ref{stabl-descrpt_tab}, Zn-{\mof} is the system with
the highest catalytic activity towards the dissociation reaction in the
{\mof} family; it also exhibits the largest reduction in CO$_2$ uptake
capacity after being exposed to humid conditions. We will thus use this
system to study the effect of water on the crystal structure.  To this
end, we exposed Zn-MOF-74 for 48 hours to three different conditions:
(i) vacuum (less than 20 mTorr) at 300~$^{\circ}$C, (ii) 8~Torr of
D$_2$O at 100~$^{\circ}$C, and (iii) 8~Torr of D$_2$O at
300~$^{\circ}$C. Figure~\ref{Zn_XRD_fig} shows the corresponding x-ray
patterns in panels b), c), and d). Panel a) serves as the reference and
shows the x-ray pattern of the pristine Zn-{\mof} before any exposure to
humid conditions. The figure clearly shows that the simple presence of
water alone does \emph{not} affect the crystal structure (c), neither
does high temperature (thermal decomposition point
$\sim$350~$^{\circ}$C\cite{Dietzel_2008:structural_changes}) (b); only
when combining the two (d) does the MOF disintegrate. It follows that
the crystal structure is only lost when the dissociation reaction takes
place, proving that the reaction products OH and H are responsible for
the degradation of Zn-MOF-74.

\begin{figure}[t]
\centering
\includegraphics[width=0.8\columnwidth]{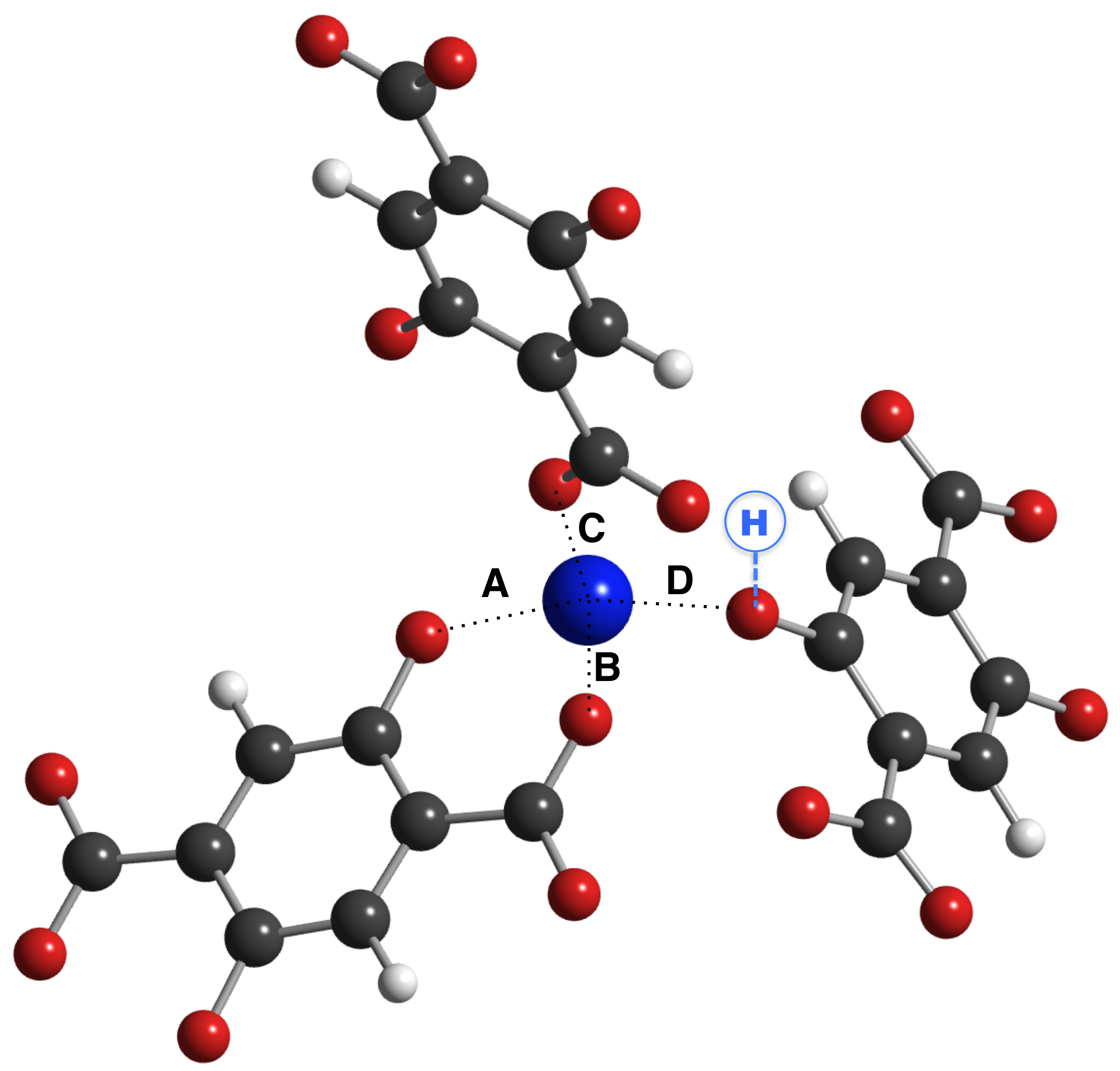}
\caption{\label{linker-names_fig}O--\metal\ bonds in \mof. Red, black,
white, and blue spheres represent O, C, H, and {\metal} atoms,
respectively. The linker making bond \bond{E} (which is very similar to
bond \bond{C}) is not shown, as it points straight out of the page.}
\end{figure}

To further support this claim and elucidate how the H$_2$O, OH, and H
groups affect the structure of the MOF-74 system, we use {\it ab initio}
calculations and investigate the weakest link in the MOF-74 structure,
{\it{i.e}}.\ the O--\metal\ bonds at the junctions between the metal
centers and the linkers. Four cases will be considered: (i) The H$_2$O
molecule is adsorbed at the metal center, (ii) the OH group is adsorbed
at the metal center and one H atom is attached to the O atom of the
linker.  Case (iii) and (iv) correspond to the scenarios of having one
OH at the metal center and one H at the O of the linker, independently.
Case (i) corresponds to a water molecule adsorbed at the metal center,
before any dissociation has occurred, whereas case (ii) correspond to
the scenario where the water molecule has split, as indicated in
Fig~\ref{1H2O_fig}.  In the MOF-74 structure, each metal center is
connected to four linkers trough their O atoms, but one linker connects
via two O atoms so that in total each metal center is bonded to five O
atoms. We will refer to these bonds as \bond{A}, \bond{B}, \bond{C},
\bond{D}, and \bond{E}, as shown in Fig.~\ref{linker-names_fig}. We
report the O--\metal\ bond length \emph{change} for cases (i)--(iv) in
Table~\ref{dimensions_tab} for Mg-, Zn-, Co-, and Ni-{\mof}.

\begin{table}[t]
\caption{\label{dimensions_tab} Calculated changes in the O--\metal\
bond length [\%] upon adsorption of H$_2$O or OH at the metal center
(\metal-site) and the attachment of H to the oxygen of the linker
(O-site, indicated in Fig.~\ref{linker-names_fig}). Results are given
for $\mathcal{M}$-MOF-74 with $\mathcal{M}=$ Mg, Zn, Co, and Ni.
Changes are reported with respect to the corresponding pristine
$\mathcal{M}$-MOF-74 systems; negative numbers indicate a shortening of
the bond.} 
\begin{tabular*}{\columnwidth}{@{}l@{\extracolsep{\fill}}ccrrrrr}\hline\hline
\metal & \metal-site & O-site &\bond{A} &\bond{B} &\bond{C} &\bond{D} &\bond{E}\\\hline
Mg   &H$_2$O   & $\times$ &0.8       &0.7       &\bf 4.2   &2.0       &1.5    \\
     &OH       & H        &$-$2.3    &1.0       &18.9      &\bf 52.1  &1.1    \\
     &OH       & $\times$ &3.1       &3.1       &\bf 16.6  &8.3       &1.7    \\
     &$\times$ & H        &7.3       &$-$2.9    &1.8       &\bf 16.5  &$-$1.5 \\\hline
Zn   &H$_2$O   & $\times$ &0.8       &3.1       &\bf 5.2   &4.1       &0.2    \\
     &OH       & H        &-4.3      &0.1       &76.9      &\bf 81.9  &4.0    \\
     &OH       & $\times$ &$-$4.0    &$-$1.0    &\bf 74.6  &67.9      &1.4    \\
     &$\times$ & H        &14.9      &13.4      &2.6       &\bf 42.5  &5.8    \\\hline
Co   &H$_2$O   & $\times$ &$-$5.4    &$-$9.4    &\bf 8.6   &0.7       &1.9    \\
     &OH       & H        &$-$3.6    &1.4       &3.2       &\bf 60.8  &3.2    \\
     &OH       & $\times$ &$-$0.3    &1.5       &2.4       &3.6       &\bf 6.3\\
     &$\times$ & H        &12.6      &$-$4.9    &5.7       &\bf 45.9  &$-$0.2 \\\hline
Ni   &H$_2$O   & $\times$ &6.0       &5.4       &$-$11.4   &\bf 8.8   &5.2    \\
     &OH       & H        &$-$2.0    &1.5       &21.5      &\bf 81.2  &$-$0.1 \\
     &OH       & $\times$ &$-$2.6    &1.3       &19.0      &\bf 71.7  &$-$0.4 \\
     &$\times$ & H        &11.6      &5.1       &--0.3     &\bf 64.6  &1.6\\\hline\hline
\end{tabular*}
\end{table}

In the first two rows for each metal, Table~\ref{dimensions_tab} shows
the bond length change caused by the adsorption of water on the metal
center and the dissociation of this molecule into OH+H. The largest
changes are highlighted in bold. Note that an excessive elongation of
bonds is the precursor for breaking. Contrary to the assumption that the
adsorption of H$_2$O on the metal centers by itself is responsible for
the structural instability of MOF-74, Table~\ref{dimensions_tab} shows
that for all cases studied (Mg-, Zn-, \mbox{CO-}, and Ni-{\mof}) H$_2$O
itself has only a negligible effect on the O--\metal\ bonds compared to
the reaction products OH+H. This shows once more that the instability of
{\mof} under humid conditions has its roots in the dissociation
reaction, at the metal sites, that produces OH and H groups, and not in
the interaction of the H$_2$O molecule and the MOF. Note that Mg- and
Co-{\mof} are the systems that exhibit the smallest elongation in the
O--\metal\ bonds, while Zn- and Ni-{\mof} are the most affected once the
{\reaction} reaction takes place.

The strong and complex effect of the reaction products OH and H combined
can be disentangled by studying their individual effects. We have
tabulated the elongation of the O--{\metal} bonds caused by these two
groups independently in Table~\ref{dimensions_tab}. It is particularly
important to note that the water dissociation OH+H at a particular metal
site, viewed from a neighboring metal side, looks like the H-only case
at the O-site. In the case where we have only an OH group at the metal
center, the O--Co bonds are the least affected, followed by O--Mg,
O--Ni, and O--Zn bonds, in that order. However, the OH group affects the
O--\metal\ bonds in different ways for each metal center and the complex
interaction between the metal centers and the OH groups does not allow
for a simple explanation of why some metals centers are more or less
affected than others. On the other hand, the attachment of only the H
group to the linker primary affects the \bond{D} bond, see
Fig.~\ref{linker-names_fig} and Table~\ref{dimensions_tab}. More
importantly, Table~\ref{dimensions_tab} shows that the elongation of
this bond follows the order O--Mg $<$ O--Zn $<$ O--Co $<$ O--Ni,
suggesting the following explanation: In the O--\metal\ covalent bond
the O pulls electrons away form the metal center. As H attaches to the O
of the linker, it donates its electron to the O. In turn, O loses its
ability of pulling electrons from the metal centers, weakening the
\bond{D} bond and ultimately elongating it. The described effect
corresponds exactly to the concept of electronegativity. For the metals
we find the following values for the Pauli
electronegativity:\cite{Allred_1961:electronegativity_values} Mg (1.31)
$<$ Zn (1.65) $<$ Co (1.88) $<$ Ni (1.91), {\it{i.e}}.\ the exact same
order found in the elongation of the \bond{D} bond. We conclude that the
electronegativity of the metal centers can be used as a descriptor of
the {\mof} stability under the effect of H groups attached to the
linker.

Mg-MOF-74, similar to Zn-MOF-74, also has a strong catalytic activity
for the water dissociation reaction---see Table~\ref{stabl-descrpt_tab}
and Fig.~\ref{linear_fig}. One might thus assume that it also easily
disintegrates in the presence of water. However, experiments confirm
that its crystal structure is unaltered after being exposed to humid
conditions,\cite{Liu_2011:stability_effects} while its CO$_2$ uptake
capacity is significantly reduced. We understand the latter in the
context of the top panel of Fig.~\ref{linear_fig}; the former can be
elegantly explained by looking at the changes in bond lengths. In
Mg-MOF-74, even though the adsorption of H and OH has a stronger effect
on the O--\metal\ bonds than the adsorption of H$_2$O, the elongation of
these bonds is significantly smaller than in Zn-, Ni-, and Co-MOF-74,
see Table~\ref{dimensions_tab}. We can thus explain the apparent
contradictory experimental finding by other
authors\cite{Liu_2011:stability_effects, Schoenecker_2012:effect_water,
DeCoste_2013:effect_water} that Mg-MOF-74 is structurally stable even
though its CO$_2$ uptake is significantly reduced in the presence of
water. Furthermore, Liu and Co-workers\cite{Liu_2011:stability_effects}
also pointed out that Ni-{\mof} looses its crystal structure after
exposing it to humid conditions, however, its CO$_2$ uptake capacity is
less affected than in Mg-MOF-74. The top panel of Fig.~\ref{linear_fig}
shows that the catalytic activity of Ni-{\mof} towards the dissociation
reaction is small, explaining why the CO$_2$ uptake capacity has a small
decrease after the exposure to water. On the other hand,
Table~\ref{dimensions_tab} shows that the H and OH groups have a large
impact on the O--Ni bonds, larger than in all the other systems,
explaining why the crystal structure of Ni-MOF-74 degrades easily under
humid conditions.

In summary, the CO$_2$ reduction uptake and loss of crystal structure
after MOF-74 is exposed to humid conditions are \emph{not} related.
Concerning the former mechanism, once MOF-74 has been exposed to water
above 150~$^{\circ}$C, the water dissociation reaction has produced the
products OH (bound to the metal center) and H (bound to the oxygen of
the linker). As the metal centers are now occupied by the OH groups,
they become poisoned, suppressing the adsorption of new guest molecules
and consequently we see a reduction in the CO$_2$ uptake capacity in all
member of the {\mof} family. This reduction is directly proportional to
the catalytic activity of the {\mof} towards the {\reaction} reaction,
which we were able to link to the radius of the metal. On the other
hand, how strong the metal centers can hold on to their linkers and
retain their structure after the water has dissociated is an entirely
different question. Concerning this latter mechanism, we find
that---depending on the electronegativity of the metal---the reaction
products OH and H by themselves may now significantly elongate the
O--\metal\ bonds and thus cause the loss of crystal structure. The two
mechanism are related only insofar as that some water needs to
dissociate in the first mechanism (responsible for the CO$_2$ uptake
reduction) so that the second mechanism can take place (loss of crystal
structure).

\subsection{Suppressing Water Dissociation to Increase MOF Structural
Stability}

\begin{figure}[t]
\centering
\includegraphics[width=0.65\columnwidth]{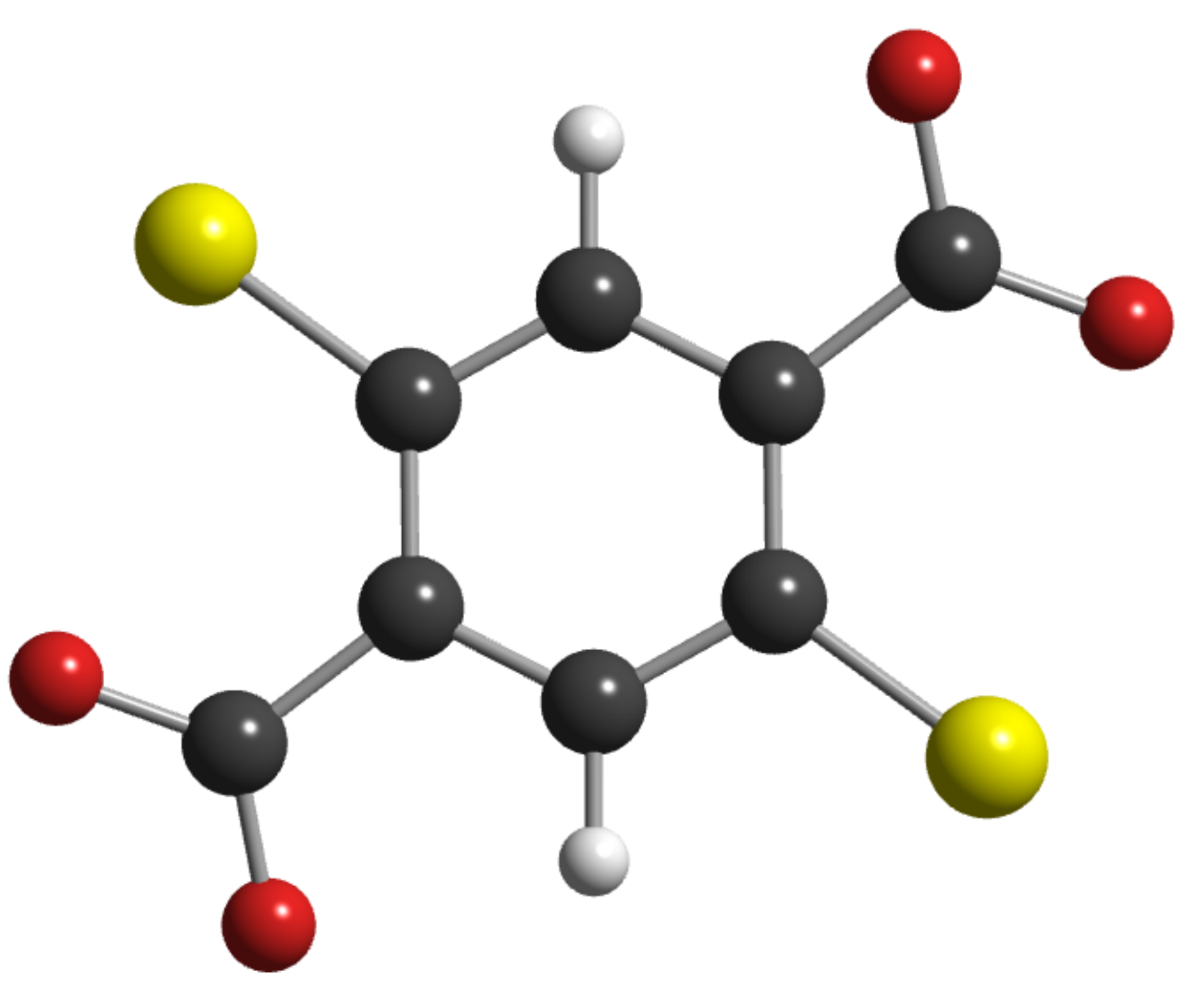}
\caption{\label{s-linker_fig} S-modified linker (H$_4$DSBDC). Black,
red, white, and yellow spheres represent C, O, H and S atoms,
respectively. The S atoms occupy the position of the O atoms in the
regular linker (H$_4$DOBDC).}
\end{figure}

\begin{figure}[t] \includegraphics[width=\columnwidth]{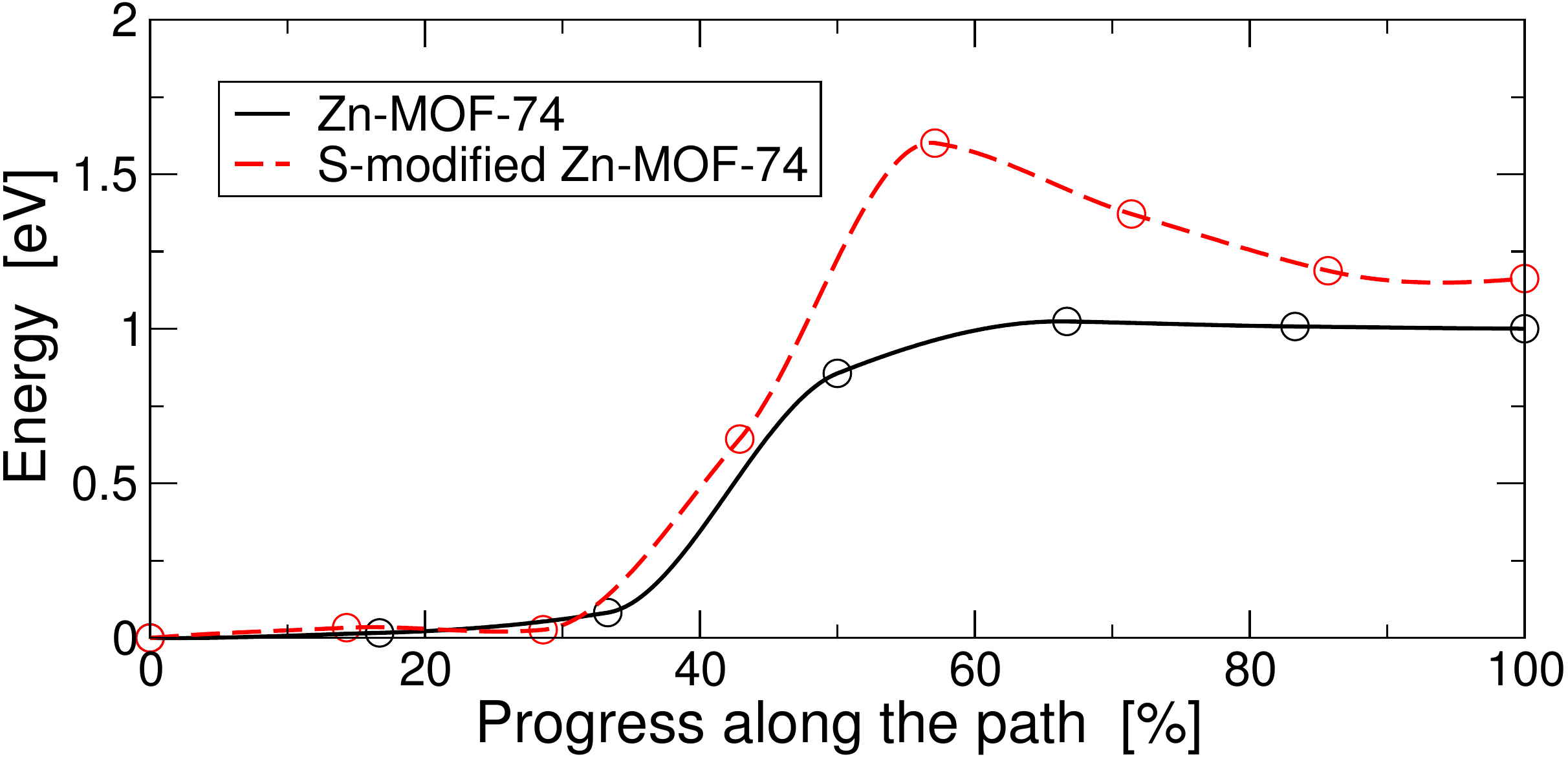}
\caption{\label{neb_fig} Activation barrier for the water dissociation
reaction H$_2$O~$\rightarrow$~OH+H in Zn-MOF-74 and S-modified
Zn-MOF-74. The barrier in the S-modified system is much higher, in
effect completely suppressing this reaction and thus stabilizing the
MOF.}
\end{figure}

With the precise knowledge of what causes the structural instability of
MOF-74, we are now in a position to propose modifications of MOF-74 that
might make it more water stable. According to our findings, the key
element is to suppress the water dissociation reaction, such that the
reaction products (OH and H) do not occur. One possible way to
accomplish this is to substitute the O atom at the linker (the one where
the H is attaching after the dissociation takes place) with a different
atom. S is a good candidate for this purpose, as it (i) has the same
number of valence electron and is thus chemically similar to O and (ii)
it is less electronegative,\cite{Allred_1961:electronegativity_values}
suggesting that it might have a less strong ``draw'' on the H from the
water at the metal site. This linker modification is shown in
Fig.~\ref{s-linker_fig}. We confirm this conjecture with \emph{ab
initio} transition-state
calculations,\cite{Henkelman_2000:climbing_image,
Henkelman_2000:improved_tangent} modeling the reaction pathway for the
H$_2$O~$\rightarrow$~OH+H reaction for the original Zn-MOF-74 and the
proposed S-modified Zn-MOF-74; results are plotted in
Fig.~\ref{neb_fig}. From this figure we see that the pristine Zn-MOF-74
has an activation barrier for the water dissociation reaction of
1.02~eV. However, the S-modified Zn-MOF-74 system has an activation
barrier of 1.60~eV, which is 0.58~eV higher. From the Arrhenius equation
we can estimate (assuming the same pre-exponential factor for both
reactions and a temperature of 200~$^{\circ}$C) that the rate of water
dissociation in S-modified Zn-MOF-74 is only $10^{-7}$ of that in
Zn-MOF-74, and thus essentially completely suppressed.

Unfortunately, we have not been able to experimentally verifiy the
stability of S-modified Zn-{\mof}. When we try to synthesize the
S-modified Zn-MOF-74 structure using the S ligand (H$_4$DSBDC), the
reactants form a different structure. We believe this is because the
formation of a strong bond between the hard acids, Zn$^{2+}$, and the
soft S is difficult. On the other hand, Sun and
co-workers\cite{Sun_2013:mn22_5-disulfhydrylbenzene-1} have been able to
synthesize S-modified Mn-MOF-74, but this MOF also eludes experimental
verification of our hypothesis, as it---with and without
modification---is extremely unstable. Our experiments showed that the
methanol exchanged S-modified Mn-MOF-74 sample decomposed quickly (< 100
min) upon contact with air, possibly due to the oxidation of the
Mn$^{2+}$ metal centers by O$_2$. This, plus the small catalytic
activity towards the {\reaction} reaction of the Mn-MOF-74 system,
prevent us from comparing the stability between the pristine and the
S-modified Mn-MOF-74. Work to synthesize and confirm the stability of
S-modified and other modifications that suppress the water dissociation
reaction in MOF-74 is currently ongoing.

\section{Summary}

In this work we offer a comprehensive explanation of the crystal
structure instability and CO$_2$ uptake reduction of MOF-74 in the
presence of water. We show that water by itself has a negligible effect
on the stability of the MOF crystal structure and no influence on the
reduction of the CO$_2$ uptake, since it can be reversibly removed from
the MOF.  We further show that the water dissociation reaction
H$_2$O~$\rightarrow$~OH+H, which is governed by the radius of the metal,
is responsible for the reduction of the CO$_2$ gas uptake and sets the
stage for the possible disintegration of the MOF.  Its products, OH and
H, might---depending on the metal and its
electronegativity---significantly elongate the O--\metal\ bonds, which
leads to the loss of the crystal structure. Future research concerning
the design of water stable MOF-74 thus needs to focus on the suppression
of the H$_2$O~$\rightarrow$~OH+H reaction at the metal centers.

\section {\label{exp_detail_sec}Methods}

\subsection{\label{CO2_uptake}CO$_2$ Uptake Measurements and
Crystal-Structure Determination}

To measure the CO$_2$ uptake of Zn-, Mg-, Ni-, and Co-MOF-74, each
MOF-74 powder ($\sim$2~mg) was pressed onto a KBr pellet ($\sim$1~cm
diameter, 1--2~mm thick). The sample was placed into a high-pressure
high-temperature cell (product number P/N 5850c, Specac Ltd, UK) at the
focal point compartment of an infrared spectrometer (Nicolet 6700,
Thermo Scientific, US). The samples were activated under vacuum at
180~${^\circ}$C for at least 4~hours; when humidity levels were reduced,
the sample was cooled down to room temperature. 6~Torr of CO$_2$ was
loaded into the cell for a period of 30~minutes, spectra were recorded
as a function of time during the adsorption process to evaluate the
CO$_2$ uptake before water exposure.  Evacuation of CO$_2$ for a period
of 30 minutes, reaching a pressure below 200~mTorr, was done prior to
the increase of temperature at 100~${^\circ}$C and 200~${^\circ}$C (two
sets of experiments); at this point, 8~Torr of heavy water (D$_2$O) was
loaded into the cell for a period of 20~minutes, spectra were recorded.
Desorption of water was done at 150~${^\circ}$C for 5~hours to assure
water removal. The sample was cooled back down to room temperature to
evaluate the CO$_2$ uptake after water exposure under the same
conditions (6~Torr of CO$_2$ for 30~minutes).

The structural instability in the presence of water was investigated in
detail for Zn-{\mof}. To this end, Zn-{\mof} pellets were fabricated and
their crystalline structure was characterized by x-ray diffraction
(Rigaku Ultima III XRD) before and after water exposure. Samples were
subjected to a pressure of 8 Torr of water at 100~$^{\circ}$C and
300~$^{\circ}$C for two days. As a control experiment, the same was done
without water at room temperature (25~$^{\circ}$C) and high temperature
(300~$^{\circ}$C).

\subsection{\label{DFT_calc}\emph{Ab Initio} Simulations}

\emph{Ab initio} calculations were performed at the DFT level with the
\textsc{Vasp} code.\cite{Kresse_1996:efficient_iterative} We used
projector augmented wave (PAW) pseudo-potentials
\cite{Kresse_1999:ultrasoft_pseudopotentials} and a plane-wave expansion
with a kinetic-energy cutoff of 600~eV. Due to the important van der
Waals interaction between the metal centers of the MOF and the guest
molecules, we used the vdW-DF exchange-correlation
functional.\cite{Thonhauser_2015:spin_signature, Berland_2015:van_waals,
Langreth_2009:density_functional, Thonhauser_2007:van_waals} Due to the
large dimensions of the unit cell, only the $\Gamma$--point was sampled.

We started from the experimental rhombohedral structure of Zn-, Mg-,
Ni-, and Co-MOF-74 with 54 atoms in its primitive cell and space group
R$\bar{3}$.\cite{Zhou_2008:enhanced_h2} Then, the positions of the atoms
and cell were relaxed until the forces were less than 1~meV/\AA.  These
systems were used to calculate the oxygen--metal (O--{\metal}) bond
elongations caused by H$_2$O, OH, and H groups. Again, in all these
cases, the atoms were relaxed until the forces were less than 1~meV/\AA.
To model the H$_2$O~$\rightarrow$~OH+H reaction we used a
transition-state search algorithm, {\it{i.e}}.\ the climbing-image
nudged-elastic band method.\cite{Henkelman_2000:climbing_image,
Henkelman_2000:improved_tangent} A graphical representation of the
MOF-74 structure can be seen in Fig~\ref{1H2O_fig}.

\section*{Acknowledgements}

This work was entirely supported by Department of Energy Grant No.\
DE--FG02--08ER46491. It further used resources of the Oak Ridge
Leadership Computing Facility at Oak Ridge National Laboratory, which is
supported by the Office of Science of the Department of Energy under
Contract DE--AC05--00OR22725.

\bibliography{references,biblio}

\end{document}